\documentstyle[12pt,a4,epsf]{article}
\newcommand{\rcite}[1]{{\cite{#1}}}
\newcommand{\rref}[1]{{(\ref{#1})}}
\newcommand{\tref}[1]{{\ref{#1}}}
\newcommand{\rlabel}[1]{{\label{#1}}}
\newcommand{\rbibitem}[1]{\bibitem{#1}}
\newcommand{\be}{\begin{equation}}
\newcommand{\ee}{\end{equation}}
\newcommand{\ba}{\begin{eqnarray}}
\newcommand{\ea}{\end{eqnarray}}

\begin{document}
\begin{titlepage}
\begin{flushright}
{NORDITA-96/29 N,P}\\
{hep-ph/yymmdd}
\end{flushright}
\vspace{2cm}
\begin{center}
{\large\bf Weak long distance contributions to the \\
neutron and proton electric dipole moments}\\
\vfill
{\bf Johan Bijnens and Elisabetta Pallante}
\\[0.5cm]
NORDITA, Blegdamsvej 17,\\
DK-2100 Copenhagen \O, Denmark\\[0.5cm]

\end{center}
\vfill
\begin{abstract}
We evaluate the long distance weak contribution to the neutron and proton
electric dipole moments using an effective Lagrangian framework. We estimate
the coefficients needed by a factorization hypothesis and additional
assumptions on $\gamma_5$ terms in the baryon lagrangian. We obtain
$|d_e^n|\approx 5\times 10^{-32}$ e$\cdot$cm and $
|d_e^p|\approx 4\times 10^{-32}$ e$\cdot$cm. 
The former estimate is similar to the quark model
estimates done previously.
\end{abstract}
\vspace*{1cm}
\vfill
 June 1996
\end{titlepage}

\section{Introduction}
\rlabel{1}
\setcounter{equation}{0}

The contributions to the electric dipole moment (EDM) of the neutron 
within the Standard model have been widely explored. In the Standard model
the main contribution comes from the strong CP-violating $\theta$-term
\rcite{Witten} and recently reconsidered in \rcite{PR}.
The experimental and theoretical situation can be found reviewed in 
\rcite{reviews}. The present experimental limits are
\be
|d_e^n| \le 11\cdot 10^{-26} e\cdot cm~\rcite{edmupper}~~ \mbox{and} 
~~d_e^p = (-3.7\pm~6.3)\cdot 10^{-23} e\cdot cm~\rcite{pdmupper}\,.
\ee
The neutron and proton EDM appear as an higher order effect of weak 
interactions. The one loop contribution with W exchange vanishes
because of KM combinations and the two-W boson loops contribution
is also vanishing as shown in \rcite{Shabalin} and references therein.
First non vanishing contributions are the so called transition quark
electric dipole moments \rcite{Shabalin,Morel,Nanopoulos} and the 
insertion of penguin diagrams \rcite{MANY} within the baryon .
Penguin diagrams can in fact produce the CP violating phases needed
to generate the EDM term.
The EDM is then generated by a two step process: the strong penguin 
diagram insertion which causes the transition $d\to s$ and weak 
radiative decay of the final strange baryon (e.g. $\Sigma^0 ,\Lambda\to 
n\gamma$). Already in \rcite{Gavela} it was observed that penguin 
diagrams' contributions dominate the EDM.
The evaluation of the long distance part of penguin insertions has been 
done up to now relying on quark models, like the one in \rcite{Gavela1}.
See \rcite{reviews} for more references.

In this letter we propose an alternative derivation of the neutron and 
proton EDM based on a factorization hypothesis which leads to the 
derivation of the EDM within the framework of chiral perturbation theory 
for baryons.

We first describe our approach and perform the calculation.
Here the assumptions made at various stages will also be explained.
Then we present
numerical results for both the proton and neutron electric dipole moment.
A comparison with power counting in the
heavy-baryon formalism for Chiral Perturbation Theory
and a proof that our contribution is the leading one are presented next.
 Finally, we recapitulate our main conclusions.

We do not attempt to ascribe an uncertainty to our results. However, contrary
to the $p$-wave hyperon nonleptonic decays we have rather small cancellations
between the different subamplitudes. We therefore expect higher orders to be
of normal size. The main uncertainty is the assumption made in estimating the
coefficients in the Lagrangian and the factorization ansatz.

\section{The calculation}

The gluonic penguin is the main source of CP violation in the weak 
$|\Delta S| =1$ hamiltonian. 
The effective interaction which mediates the 
$d\to s$ transition in the EDM diagram involves the strong penguin
four-quark operators of the effective weak $|\Delta S|=1$
Lagrangian ${\cal L}_{eff}=-G_F/\sqrt{2}~\sum_{i=3}^6~C_i(\mu )Q_i(\mu 
)$ (we use the definitions of the $Q_i$ as in \rcite{BURAS}).
The operator $Q_6$ is defined as 
$Q_6=(\bar{s}_\alpha d_\beta )_{V-A}\sum_q(\bar{q}_\beta 
q_\alpha )_{V+A}$, where $V\pm A$ stands for the combination $\gamma_\mu 
(1\pm\gamma_5)$ and $\alpha , \beta$ denote colour indices.
Using Fierz identities one can rewrite $Q_6$ as follows
\be
Q_6=-8\sum_q~\bar{s}_{\alpha_L} q_{\alpha_R}~\bar{q}_{\beta_R} d_{\beta_L}
+4\sum_q~\bar{s}_{\alpha_R}\gamma_\mu q_{\alpha_R}~\bar{q}_{\beta_L}\gamma_\mu
 d_{\beta_L}\, ,
\ee
where $q_{R,L}=(1\pm\gamma_5)/2~q$.
At this point we introduce a factorization hypothesis:\be
\langle B_i |Q_6|B_j\rangle=
-8\langle\bar{d}_R d_L\rangle\langle B_i|\bar{s}_L d_R|B_j\rangle\,.
\rlabel{LQQ}
\ee
Within the factorization hypothesis all the other operators $Q_i,~i=3,4,5$
do not contribute. The hypothesis is favoured by the substantial 
enhancement of the coefficient of the operator $Q_6$ by next-to-leading 
corrections and the enhancement of the $Q_6$ contribution to weak non 
leptonic decays.
For later use we introduce $\bar{q}\lambda_{\pm}q$ with 
$\lambda_{\pm}$ projection 
matrices defined in terms of the Gell-Mann matrices 
$\lambda_6, \lambda_7$ as 
$\lambda_{\pm}=(\lambda_6\pm i\lambda_7)/2$.

The baryon Lagrangian for strong interactions in presence of external 
scalar and pseudoscalar sources allows for the general form leading in the 
derivative expansion and in the light quark mass matrix: 
\ba
{\cal L}_B&=& Tr\left(\bar{B}(iD_\mu\gamma^\mu-m_B)B\right) \nonumber\\
&&+b_1Tr\left(\bar{B}\chi_+B\right) +b_2Tr\left(\bar{B}B\chi_+\right) 
+b_3 Tr\left(\bar{B}B\right)~Tr\left(\chi_+\right)
\nonumber\\
&&+b_1^5Tr\left(\bar{B}\chi_-\gamma_5 B\right) 
+b_2^5 Tr\left(\bar{B}\gamma_5 B\chi_-\right)
+b_3^5 Tr\left(\bar{B}\gamma_5 B \right)~Tr\left(\chi_-\right) ,
\rlabel{LBB}
\ea
where $Tr$ stands for the trace over flavour indices and the covariant 
derivative in the case of interest contains the electromagnetic
field $D_\mu B=\partial_\mu B +ie{\cal A}_\mu [Q,B]$. The field 
$\chi_\pm = \xi^\dagger\chi\xi^\dagger \pm\xi\chi^\dagger\xi$ contains
the external scalar and pseudoscalar sources with 
$\chi =2B_0(s(x)+ip(x))$ and, in the absence of meson field ($\xi =1$),
we have $\chi_+=4B_0s(x)$ and $\chi_-=4iB_0p(x)$.
$B_0$ is related to the scalar quark condensate through the 
identity $\langle 0|\bar{q}q| 0\rangle_{q=u,d,s}=-B_0(f^2/2)(1+O({\cal M}))$,
with $f\simeq f_\pi\simeq 132$ MeV and ${\cal M}$ the light quark mass 
matrix.
The baryon fields are incorporated in the $3\times 3$ matrix
\[ 
B= \left( 
\begin{array}{ccc}
{\Sigma^0\over \sqrt{2}}+{\Lambda\over \sqrt{6}} & \Sigma^+ & p \\
\Sigma^- &  -{\Sigma^0\over \sqrt{2}}+{\Lambda\over \sqrt{6}} & n \\
\Xi^- & \Xi^0 & -\sqrt{2\over 3}\Lambda 
\end{array}  \right) 
\]
and transform non linearly under $SU(3)_L\times SU(3)_R$.
The electromagnetic coupling of the baryon field in the covariant 
derivative generates the ordinary magnetic moment for baryons.
 
Taking the functional derivative 
with respect to $s_{ij},p_{ij}$ of the generating 
functional of the baryon Lagrangian including terms \rref{LBB}  
it follows that
\ba
\langle B_k|\bar{q}_iq_j| B_l\rangle&=&
\langle B_k|-4B_0b_1\bar{B}_iB_j-4B_0b_2(\bar{B}B)_{ji}-4B_0b_3
Tr\bar{B}B
\delta_{ij}|B_l\rangle \\
\langle B_k|\bar{q}_i\gamma_5q_j| B_l\rangle&=&
\langle B_k|4B_0b_1^5\bar{B}_i\gamma_5B_j+4B_0b_2^5
(\bar{B}\gamma_5B)_{ji}+4B_0b_3^5 Tr\bar{B}\gamma_5B
\delta_{ij}|B_l\rangle\, .
\nonumber
\ea
Both identities relate the two fermion matrix element \rref{LQQ} to the 
corresponding baryon Lagrangian. 
In the case of the flavour combination $\bar{B}_iB_j$ with $i,j=s,d$
also the ordinary baryon Lagrangian
for $|\Delta S|=1$ weak interactions has to be taken into account. 
Assuming octet enhancement it has two terms at leading order in the
derivative expansion which transform as $(8_L,1_R)$:
\be
{\cal L}_{eff}^{|\Delta S| =1}= 
a~Tr\bar{B}\{\lambda_6, B\} +b~Tr\bar{B}[\lambda_6,B] ,
\rlabel{LW}
\ee
where for the case of interest the meson field is absent, i.e. $\xi =1$.
Then the full baryon Lagrangian which induces $s\to d$ transitions
can be written as follows
\ba
{\cal L}_B^{s\to d}=&&4B_0 \biggl ( {\alpha+\gamma\over 2}b_1
Tr\bar{B}\lambda_-B
+{\alpha+\delta\over 2}b_2Tr\bar{B}B\lambda_-\biggr ) \nonumber\\
&&+4B_0 \biggl ( {\alpha^\star +\gamma\over 2}b_1
Tr\bar{B}\lambda_+B
+{\alpha^\star +\delta\over 2}b_2Tr\bar{B}B\lambda_+\biggr ) \nonumber\\
&&-4B_0 {\alpha\over 2}\biggl ( b_1^5
Tr\bar{B}\lambda_-\gamma_5B+b_2^5Tr\bar{B}\gamma_5B\lambda_-\biggr ) 
\nonumber\\
&&+4B_0 {\alpha^\star\over 2}\biggl ( b_1^5
Tr\bar{B}\lambda_+\gamma_5B+b_2^5Tr\bar{B}\gamma_5B\lambda_+\biggr ) . 
\rlabel{LBF}
\ea
The parameters $\gamma,\delta$ are
defined as $2B_0\gamma b_1=a+b$ and
$2B_0\delta b_2=a-b$, where $a,b$ are the couplings in \rref{LW}.
The Lagrangian \rref{LBF} induces transitions $n\to \Sigma^0,\Lambda$
and $p\to \Sigma^+$ both parity conserving (p.c.) and 
parity violating (p.v.) i.e. with a $\gamma_5$ insertion. 
Here we have made the assumption that the parity violating part of the other
operators besides $Q_6$ can be neglected.
The EDM term of the neutron ${\cal L}_{EDM}=i{d_e\over 
2}\bar{\psi}\sigma_{\mu\nu}F^{\mu\nu}\gamma_5\psi$ can be generated at tree 
level in the baryon theory by the vertices in \rref{LBF} and through  
the insertion of the anomalous magnetic moment operator which we write as
\be
{\cal L}_\mu ={e\over 4m_N}\biggl (
\mu_DTr\bar{B}\sigma_{\mu\nu}F^{\mu\nu}\{ Q,B\}
+\Delta\mu_F Tr\bar{B}\sigma_{\mu\nu}F^{\mu\nu}[Q,B]\biggr ) .
\rlabel{LMAG}
\ee
The baryon magnetic moments receive the ordinary
contribution from the leading 
electromagnetic coupling in \rref{LBB} and the anomalous contribution
from the next to leading $O(p^2)$ terms in \rref{LMAG}.

\begin{figure}
\begin{center}
\leavevmode\epsfxsize=14cm\epsfysize=6cm\epsfbox{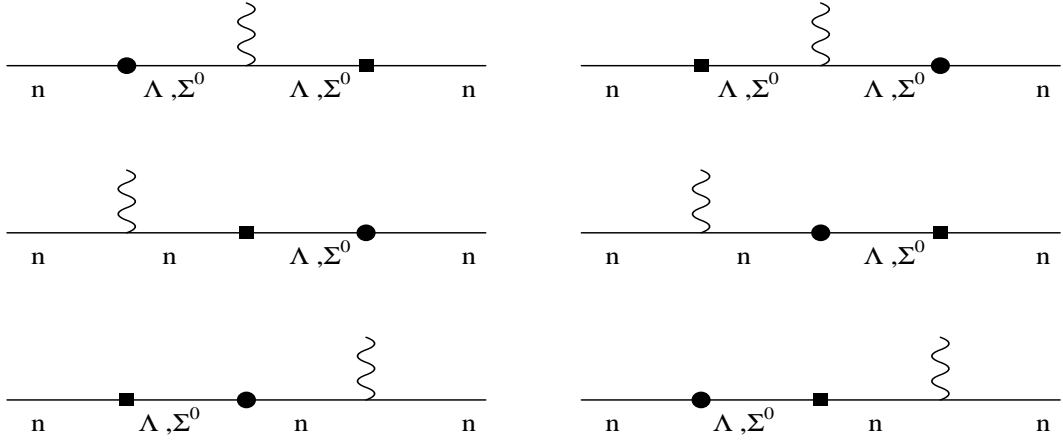}
\end{center}
\caption{Tree level diagrams which contribute to the neutron electric dipole
moment. The circle vertex is the parity conserving vertex. The box is the 
parity violating vertex. The photon insertion is the anomalous magnetic moment
vertex. In the proton case the same diagrams contribute where the 
$p\to \Sigma^+$ transition is allowed.}
\label{edmfig1}
\end{figure}

At tree level in the full baryon theory the
set of diagrams which contribute to the electric dipole moment of 
the neutron and proton are shown in Fig.\tref{edmfig1}. 
They are given by the insertion of a parity violating (p.v.) vertex and 
a parity conserving (p.c.) vertex from \rref{LBF} and the insertion
of the anomalous magnetic moment vertex in \rref{LMAG}.

The Lagrangians \rref{LBF} and \rref{LMAG} lead to the 
following tree level contribution for the neutron case:
\ba
d_e^n&=& -{ie\over 2m_N}{\alpha -\alpha^*\over 2}\biggl\{
\mu_\Lambda A_\Lambda A_\Lambda^5 {1\over M_\Lambda^2-M_n^2}\nonumber\\
&&+\mu_{\Sigma^0} A_\Sigma A_\Sigma^5 {1\over M_{\Sigma^0}^2-M_n^2}
\nonumber\\
&&+\mu_n A_\Lambda A_\Lambda^5 {M_\Lambda\over M_n(M_\Lambda^2-M_n^2)}
\nonumber\\
&&+\mu_n A_\Sigma A_\Sigma^5 {M_{\Sigma^0}\over M_n(M_{\Sigma^0}^2-M_n^2)}
\nonumber\\
&&+\mu_{\Lambda\Sigma^0} A_\Lambda A_\Sigma^5 {1\over (M_\Lambda -M_n)
(M_{\Sigma^0}+M_n)}
\nonumber\\
&&+\mu_{\Lambda\Sigma^0} A_\Sigma A_\Lambda^5 {1\over (M_{\Sigma^0} -M_n)
(M_\Lambda+M_n)}\biggr \} .
\rlabel{EDMN}
\ea
The last two terms are the mixed $\Sigma^0-\Lambda$ exchange contributions.
The $\mu_i$ are the magnetic moments of the neutral baryons 
in units of nuclear magnetons (1 nuclear magneton=$e/2m_N=1.052\cdot 
10^{-14}$ e$\cdot$cm) and we 
assumed as valid their tree level expressions in terms of $\mu_D$:

\be
\mu_n=-{2\over 3}\mu_D~~~\mu_\Lambda =-{1\over 3}\mu_D~~~\mu_{\Sigma^0}=
{1\over 3}\mu_D~~~\mu_{\Lambda\Sigma^0}={1\over\sqrt{3}}\mu_D,
\ee
while the parameters $A_i$ are a short hand notation for
\ba
A_\Sigma &=&-\sqrt{2}B_0\delta b_2~~~~~~
A_\Sigma^5=2\sqrt{2}B_0b_2^5 \nonumber\\
A_\Lambda &=&\sqrt{{2\over 3}}B_0(\delta b_2-2\gamma b_1)
~~~A_\Lambda^5=-2\sqrt{{2\over 3}}B_0( b_2^5-2b_1^5)\nonumber\\
\rlabel{AIVAL}
\ea
and are related to the weak non leptonic hyperon decay 
amplitudes $\Sigma^-\to n\pi^-$, 
$\Lambda\to p\pi^-$ as it is explained in the next section.
Note that the anomalous magnetic moment term with coupling $\mu_D$ is
the only contribution to the magnetic moments of neutral baryons.

In the proton case the only possible transition induced by \rref{LBF} is
$p\to\Sigma^+$. It involves only $b_2,b_2^5$ type of coefficients.
In the charged case also the anomalous magnetic moment term proportional 
to $\Delta\mu_F$ does contribute. 
The expression for the electric dipole moment of the proton induced by 
Lagrangians \rref{LBF} and \rref{LMAG} is 
\be
d_e^p=-{ie\over 2m_N}(\alpha-\alpha^*) {A_\Sigma A_\Sigma^5\over 
M_{\Sigma^+}^2-M_p^2}\biggl (\Delta\mu_{\Sigma^+} +\Delta\mu_p 
{M_{\Sigma^+}\over M_p}\biggr ) .
\rlabel{EDMP}
\ee 
Here $\Delta\mu_p=\Delta\mu_{\Sigma^+}=\mu_p-1=1/3\mu_D+\Delta\mu_F$ 
are the anomalous magnetic moments of the proton and the $\Sigma^+$ 
in units of nuclear magnetons which are equal in the $SU(3)$ limit. 
They receive contributions from $\Delta\mu_F$ and $\mu_D$.
The proton EDM involves only the anomalous magnetic moment terms.
It is easy to verify that the tree level diagrams as in 
Fig.\tref{edmfig1} where the anomalous magnetic moment vertex is 
replaced by the ordinary electromagnetic coupling in the proton case
sum to zero.

The expressions \rref{EDMN} and \rref{EDMP} are the tree level contributions
to the electric dipole moment of the neutron and the proton respectively.
In section \tref{loops} we explicitly show the power counting for the tree
level and quantum corrections in the Heavy Baryon Chpt and that one loop
corrections are naturally suppressed also in virtue of the absence of large 
cancellations at tree level.

\section{Numerical results}

The numerical estimate of the tree level contributions to the neutron 
electric dipole moment in \rref{EDMN} and the proton one in \rref{EDMP}
requires the knowledge of the following set of weak and strong parameters:
the combination $\alpha-\alpha^*$, the parameters 
$A_{\Lambda ,\Sigma}$, $A_{\Lambda ,\Sigma}^5$ and finally 
the magnetic moment coefficients $\mu_D,\Delta\mu_F$.

The latter can be extracted at tree level from the measured values
of the magnetic moments of baryons. At tree level the $SU(3)$ symmetric
Coleman-Glashow relations amongst magnetic moments are valid, while they are 
experimentally violated by about 0.25 nuclear magnetons in average 
\rcite{JMMAG}.
For the leading quantum corrections to the 
magnetic moment of baryons in the Heavy Baryon expansion see e.g.
\rcite{JMMAG}. 

If we use the experimental values of $\mu_p$ and $\mu_n$
to determine $\mu_D$ and $\Delta\mu_F$ and disregard quantum corrections
we obtain $\mu_D=-{3\over 2}\mu_n=2.87$ and 
$\Delta\mu_F=\mu_p+{1\over 2}\mu_n-1=0.8365$ for the experimental values
$\mu_p=2.793$, $\mu_n=-1.913$ in units of nuclear magnetons.

The other magnetic moments in the tree level approximation are:
\ba
&& \mu_\Lambda={1\over 2}\mu_n ~~~~~~~~~~~(-0.613\pm 0.004=-0.96)\nonumber\\
&&\mu_{\Sigma^0}=-{1\over 2}\mu_n \nonumber\\
&&\mu_{\Lambda\Sigma^0}=-{\sqrt{3}\over 2}\mu_n~~~~ 
(\pm1.61\pm 0.08=1.66)\nonumber\\
&& \mu_{\Sigma^+}=\mu_p~~~~~~~~~~~~(2.458\pm 0.010=2.793).
\ea 
In brackets the latest experimental values are indicated \rcite{RPP} and 
compared with the $SU(3)$ symmetric value. The $\Sigma^0$ magnetic 
moment has not been measured.
Even though the observed magnetic moments do not satisfy the $SU(3)$ relations
very well, a more accurate
treatment is unnecessary  in view of the other uncertainties involved.

The complete determination of the $b_i, b_i^5$ parameters requires an 
additional assumption to relate the experimentally constrained $b_i$ to
the unconstrained $b_i^5$. We impose $b_i^5=b_i$ for $i=1,2$ ($b_3$, 
$b_3^5$ do not enter the EDM expression). This choice 
seems natural starting from the Lagrangian \rref{LBB} and is our second main
assumption.

The linear combinations $B_0b_1m_s$ and $B_0b_2m_s$ enter the mass terms of the 
baryons as implied by \rref{LBB}. 
Defining $\bar{m}=(m_u+m_d)/2$ we use the combinations of baryon masses
which are not affected by the isospin breaking effect at tree level.
They are
\ba
m_N&\equiv&{M_p+M_n\over 2}=m-4B_0b_1\bar{m}-4B_0b_2m_s
~~~(=938.91897(28)~MeV)\nonumber\\
M_{\Sigma^0}&=&{M_{\Sigma^+}+M_{\Sigma^-}\over 2}=
 m -4B_0(b_1+b_2)\bar{m}~~(1192.55(8)=1193.41(5)~MeV)
\nonumber\\
M_{\Xi}&\equiv&{M_{\Xi^0}+M_{\Xi^-}\over 2}=
m-4B_0b_1m_s-4B_0b_2\bar{m}~~(=1318.07(11))~MeV)\nonumber\\
M_\Lambda&=&m-{4\over 3}B_0(b_1+b_2)(\bar{m}+2m_s)~~(=1115.57(6)~MeV),
\rlabel{NOISO}
\ea
where $m=m_B-4B_0b_3(2\bar{m}+m_s)$ takes into account the contribution 
from the $b_3$ term in \rref{LBB}. The values in brackets are the 
latest experimental determinations \rcite{RPP}.
Using the experimental values of the four masses $m_N, M_{\Sigma^0},
M_\Xi , M_\Lambda$ in \rref{NOISO}
we can  determine the combinations $B_0b_1$ and $B_0b_2$ with fixed 
$\bar{m}$ and $m_s$ (or alternatively the combinations $B_0b_1m_s$ and
$B_0b_2m_s$ if we approximate $\bar{m}=0$).
Using the set $(M_{\Sigma^0} ,m_N, M_\Xi )$ we get
\ba
M_{\Sigma^0}-m_N&=&4B_0b_2(m_s-\bar{m})\simeq 253.63\nonumber\\
M_{\Sigma^0}-M_\Xi&=&4B_0b_1(m_s-\bar{m})\simeq -125.52\nonumber\\
\rlabel{MASSDIF}
\ea
and with $\bar{m}=6$ MeV, $m_s=175$ MeV we get $2B_0b_1=-0.3714$ and 
$2B_0b_2=0.7504$. These values give $M_\Lambda\simeq 1107.15$ MeV, with 
$m$ extracted from $M_{\Sigma^0}$, which 
is a reasonable approximation of the real value. Alternatively
if we use the set $(M_{\Sigma^0} ,m_N, M_\Lambda )$ the numbers change 
to $2B_0 b_1=-0.4088$ and $2B_0 b_2=0.7504$ and a slightly too small value
for the $\Xi$ mass $M_\Xi=1054.38$.

We still need an additional constraint to fix $b_1, b_2$ together with the 
coefficients of the weak Lagrangian $\gamma,\delta$,
or equivalently $a$ and $b$. 
The latter enter the weak non leptonic
hyperon decay amplitudes. There are seven measurable amplitudes:
$\Sigma^{\pm}\to n\pi^\pm$, $\Sigma^+\to p\pi^0$, $\Lambda\to n\pi^0$, 
$\Lambda\to p\pi^-$, $\Xi^-\to\Lambda\pi^-$, $\Xi^0\to\Lambda\pi^0$
and three isospin relations both for S-wave and P-wave amplitudes.
For the chosen four independent S-wave amplitudes at tree level one 
has (for the standard definition of the S and P-wave amplitudes in weak
hyperon decays see e.g. \rcite{HYPERON1}):
\ba
A^{(S)}(\Sigma^+\to n\pi^+ )&=&0~~~~~~~~~~~~(0.06\pm 0.01)\nonumber\\
A^{(S)}(\Sigma^-\to n\pi^- )&=&{b-a\over f}~~~~~~(1.88\pm 0.01)\nonumber\\
A^{(S)}(\Lambda\to p\pi^- )&=&{a+3b\over \sqrt{6}f}~~~~(1.42\pm 0.01)
\nonumber\\
A^{(S)}(\Xi^-\to \Lambda\pi^- )&=&{a-3b\over \sqrt{6}f}~~~~(-1.98\pm 
0.01),
\ea
where the 
last number in parenthesis on the r.h.s. is the corresponding 
experimental value in units of $G_Fm_{\pi^+}^2$
(this is in agreement with ref. \rcite{HYPERON2}
since experimental values for the decay parameters of hyperon non 
leptonic decays are unchanged since the Particle Data Book of 1990
\rcite{RPPold}).
We use the values $a=-0.58\pm 0.21$ and $b=1.40\pm 0.12$ in units of 
$G_Fm_{\pi^+}^2 f_\pi$ ($f_\pi\simeq$ 132 MeV)
\rcite{HYPERON2} determined with a tree level least squares fit of 
the seven measured S-wave amplitudes. These determine the combinations
\ba
2B_0\gamma b_1&\equiv&a+b=0.82~ G_Fm_{\pi^+}^2f_\pi\nonumber\\
2B_0\delta b_2&\equiv&a-b=-1.98 ~ G_Fm_{\pi^+}^2f_\pi ,
\rlabel{HYPVAL}
\ea
with $G_F m_{\pi^+}^2 f_\pi = 2.98\cdot 10^{-8}$ GeV. 
Using instead only the measured values of the decay amplitudes of the
two processes $\Sigma^-\to n\pi^-$ and 
$\Lambda\to p\pi^-$ for a tree level determination of $a$ and $b$ 
one gets: $a=-0.54$ and $b=1.34$.

The combinations \rref{HYPVAL} determine the values of the 
parameters $A_i$ defined in \rref{AIVAL}. We obtain
\ba
A_\Sigma&=&-\sqrt{2}B_0\delta b_2\simeq 1.40~ 
G_F m_{\pi^+}^2 f_\pi
\nonumber\\
A_\Lambda&=&\sqrt{{2\over 3}}B_0(\delta b_2-2\gamma b_1)\simeq -1.48~ G_F 
m_{\pi^+}^2 f_\pi ,
\ea
while the numerical values obtained by using directly the experimental
values for $\Sigma^-\to n\pi^-$ and $\Lambda\to p\pi^-$ decays are 
$1.33~G_F m_{\pi^+}^2 f_\pi$ and $-1.42~G_F m_{\pi^+}^2 f_\pi$ respectively.
Using then $b_i=b_i^5$ and the values in \rref{MASSDIF} we also get 
\ba
A_\Sigma^5&=&2\sqrt{2}B_0b_2^5={M_{\Sigma^0} -m_N\over \sqrt{2}
(m_s-\bar{m})}\simeq 1.06
\nonumber\\
A_\Lambda^5&=&-2\sqrt{{2\over 3}}B_0(b_2^5-2b_1^5)=
-{2M_\Xi-M_{\Sigma^0}-m_N\over\sqrt{6}(m_s-\bar{m})}\simeq -1.22,
\ea
while for $A_\Lambda^5$ one gets $-1.28$ if using the experimental values
of $m_N, m_\Sigma$ and  $m_\Lambda$. 

The last parameter to be estimated is $\alpha -\alpha^*$. In terms of the
Wilson coefficient function of the effective four-quark operator $Q_6$ we have
\be
{\alpha -\alpha^*\over 2}=-8i\langle\bar{d}_Rd_L\rangle
~{G_F\over\sqrt{2}}~Im~ C_6,
\ee
where 
\be
C_6(\mu )=V_{ud}V_{us}^*z_6(\mu )-V_{td}V_{ts}^*y_6(\mu )
\ee
and
\be
Im~C_6(\mu )=-Im~V_{td}V_{ts}^*~y_6(\mu )
=c_{23}s_{23}s_{13}\sin\delta_{13}~y_6(\mu ).
\ee
The estimate of the size of the coefficient $y_6(\mu )$ is affected
by large uncertainty. We use the 
renormalization scheme independent definition in \rcite{BURAS} where the 
next-to-leading corrections at a given $\mu$ to the effective 
hamiltonian are shifted into the Wilson coefficient functions.
As noticed there the 
coefficient $y_6(\mu )$ is a very 
sensitive function of $\Lambda_{\bar{MS}}$ and $\mu$ being 
next-to-leading corrections sizable.
We use the approximate value $y_6\simeq -0.13$ at $\mu\simeq 1$ GeV and 
$\Lambda_{\bar{MS}}\simeq 300$ MeV.
The CKM matrix elements are \rcite{RPP}
$s_{23}=\vert V_{cb}\vert =0.040\pm 0.005$,
$s_{13}=\vert V_{ub}\vert \simeq0.0032$ extracted from $\vert 
V_{ub}/V_{cb}\vert =0.08\pm 0.02$. We approximate cosines to unity
and put $\sin\delta_{13}\simeq 1$ \rcite{SINDEL}. 
Using for the scalar quark condensate
$\langle\bar{d}_Rd_L\rangle =-1/2\cdot (0.235)^3~GeV^3$, this gives 
$(\alpha -\alpha^*)/ 2 =-i~5.48\times 10^{-12}$ GeV.

For the final prediction we use $A_\Sigma =1.40~G_Fm_{\pi^+}^2f_\pi$, 
$A_\Lambda =-1.48~G_Fm_{\pi^+}^2f_\pi$, $G_Fm_{\pi^+}^2f_\pi=3.0\times 
10^{-8}$ GeV and the experimental values for the magnetic moments
and baryon masses in the EDM formulas \rref{EDMN} and \rref{EDMP}.
We use $\mu_{\Lambda\Sigma^0}=+1.61$.
We predict the following value for the neutron EDM:
\ba
d_e^n= -{ie\over 2m_N}{\alpha -\alpha^*\over 2}\biggl\{&&-9.18\times 
10^{-8}~GeV^{-1} 
 \nonumber\\
&&+7.89\times 10^{-8}~GeV^{-1} 
\nonumber\\
&&-34.02\times 10^{-8}~GeV^{-1} 
\nonumber\\
&&-20.04\times 10^{-8}~GeV^{-1} 
\nonumber\\
&&-20.19\times 10^{-8}~GeV^{-1} 
\nonumber\\
&&-15.87\times 10^{-8}~GeV^{-1} 
\biggr \} .
\rlabel{EDMNNUM}
\ea
where the numerical value in each line corresponds to the relative 
expression in \rref{EDMN}. This shows the absence of large 
cancellations. This gives
\be
d_e^n\approx 5.3\times 10^{-32}~e\cdot cm
\ee
For the proton we have:
\be
d_e^p\approx -3.6\times 10^{-32}~e\cdot cm,
\ee
where we used the experimental values for $\Delta\mu_{\Sigma^+}=1.458$ 
in units of nuclear magnetons, $M_{\Sigma^+}=1189.37(6)$ MeV.
Both the neutron and proton electric dipole moments acquire the opposite 
sign if we use instead $b_i^5=-b_i$.

\section{Power counting and loops}
\rlabel{loops}

The purpose of this section is to derive the power counting rules 
for quantum corrections to the tree level electric dipole moment
contributions. They can be consistently derived within the Heavy Baryon
chiral perturbation expansion (HBChPt).

One comment is in order concerning the heavy baryon mass limit
of our tree level 
contribution to the electric dipole moment term.
In the ordinary Heavy Baryon ChPt the EDM term is one of the possible 
counterterms which appears in the tree level Lagrangian at order $p^2$ in 
the derivative expansion.
As an example the first tree level diagram shown in Fig.\tref{edmfig1} 
of the full baryon theory
leads to the following contributions in the heavy baryon mass limit
\be
-2i{\bar{B}_f^v (v^\mu S_v^\nu-v^\nu S_v^\mu )F_{\mu\nu}B_i^v
\over m_i^2-m_{pole}^2}
+2i{\bar{B}_f^v (k^\mu S_v^\nu-k^\nu S_v^\mu )F_{\mu\nu}B_i^v
\over (m_i^2-m_{pole}^2)(m_i-m_{pole})} .
\rlabel{HBlimit}
\ee
These terms are enhanced by one inverse power of the baryon 
mass splitting respect to the ordinary EDM counterterm appearing at order 
$p^2$ in the derivative expansion.

The heavy baryon mass limit in \rref{HBlimit} of the tree level 
contribution to the electric dipole moment
derived in the full theory shows that the leading
term appears at order $p$ in the derivative expansion, while the usual
first counterterm to the electric dipole moment appears at order $p^2$
both in the full theory and in the HBChPt.
This power counting for \rref{HBlimit} is consistent with the fact
that baryon propagators count as $1/p$ in the derivative expansion and
that the leading parity violating vertex in the HBChPt appears
at order $p$ i.e. $\bar{B}_v\gamma_5\gamma_\mu D^\mu B_v\to
-2\bar{B}_vS_v^\mu D_\mu B_v$.
In this case the octet mass splitting, proportional to the off-shellness
of the baryon propagator, counts as $O(p)$, while in the strange quark 
mass expansion it is $O(m_s)$. 

The full baryon Lagrangian contributing to the one loop corrections to the 
electric dipole moment term includes: the Lagrangian which mediates
$|\Delta S|=1$ weak interactions, given by \rref{LW} with the inclusion of 
the meson field through the substitution $\lambda_6\to 
\xi^\dagger\lambda_6\xi$, the usual strong interaction Lagrangian 
with the inclusion of meson interactions starting at order $p$, 
the magnetic moment term at order $p^2$, the electric dipole moment 
counterterm at order $p^2$ and the Lagrangian \rref{LBF}
with the inclusion of the meson field through the substitution 
$\lambda_-\to \xi\lambda_-\xi , \lambda_+\to 
\xi^\dagger\lambda_+\xi^\dagger$.
Both octet and decuplet states can contribute inside the loop. 

\begin{figure}
\begin{center}
\leavevmode\epsfxsize=15cm\epsfysize=3cm\epsfbox{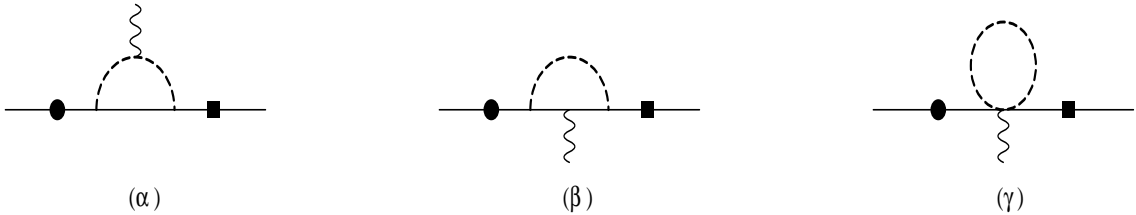}
\end{center}
\caption{One loop diagrams of class a) which contribute to the 
neutron and proton electric dipole moments. 
The circle vertex is the parity conserving (p.c.) vertex. The box is the 
parity violating (p.v.) vertex. The photon insertion in $(\beta )$ 
and $(\gamma )$ is the 
magnetic moment vertex. Internal baryon propagators can be also 
decuplet.}
\label{edmfig2}
\end{figure}

One loop diagrams contributing to the EDM term
can be divided into four classes:  a) corrections to the one loop 
contribution to the magnetic moment through the
insertion of the p.v. and p.c. vertices of \rref{LBF}, 
b) one loop correction to the p.c. vertex and to the p.v. 
vertex in \rref{LBF}, c) one loop corrections to the EDM 
vertex appearing at order $p^2$, d) one-loop diagrams with the meson
loop bridging several
of the p.v., p.c. and magnetic moment vertices, including the case where
the p.v. and/or the p.c. vertices emit the meson line. Diagrams with the 
insertion of a photon-meson-baryon-baryon vertex do not contribute.
The first class is shown in Fig.\tref{edmfig2} and the one loop
corrections to the magnetic moment of baryons in the HBChPt have been
derived in \rcite{JMMAG}.

\begin{figure}
\begin{center}
\leavevmode\epsfxsize=13cm\epsfysize=2cm\epsfbox{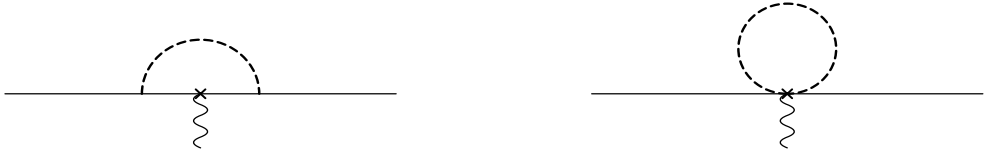}
\end{center}
\caption{One loop diagrams of class c) which contribute to the 
neutron and proton electric dipole moments. 
 The photon insertion is always the EDM counterterm that appears at
order $p^2$ in the chiral perturbation expansion}
\label{edmfig3}
\end{figure}

The magnetic moment one loop contribution is of order $p^3$ in diagram
$(\alpha )$ and of order $p^4$ in diagrams $(\beta )$ and $(\gamma )$. 
So the full contribution
to the nucleon EDM is of order $p^2$ in diagram $(\alpha )$ and of order 
$p^3$ in diagrams $(\beta )$ and $(\gamma )$. 
In the strange quark mass expansion the counting is somewhat anomalous 
because the leading tree level contribution starts at order $1/m_s$.
One loop diagrams give non analytic corrections in the strange quark 
mass. The one loop in diagram $(\alpha )$ gives a correction 
$\sqrt{m_s}$ to the
tree level diagram, while the one loop in diagrams 
$(\beta )$ and $(\gamma )$ gives a $m_s\ln m_s$ correction.

Diagrams of class b) have the same counting as the corresponding 
diagrams of class a). They include also the electric charge vertex 
insertion in the proton case.
Diagrams of class c) start at order $p^4$ and are shown in 
Fig.\tref{edmfig3}.
Diagrams of class d) have the same counting as those of a) and only
appear at order $p^2$.

This shows that within the chiral perturbation expansion the tree level
contributions to the electric dipole moment
are in fact the leading contributions.
One loop corrections are suppressed both in the derivative 
and strange quark mass expansion.

\section{Conclusions}
In this letter we have provided a new way of deriving the long distance weak
contribution to the proton and neutron electric dipole moments. 
Our final results are
\ba
d_e^n &\approx&\pm 5.3\times 10^{-32}~e\cdot cm \nonumber\\
d_e^p &\approx&\mp 3.6\times 10^{-32}~e\cdot cm  \, .
\ea
The opposite sign is a consequence of the opposite sign in the relevant
anomalous magnetic moments.
These numbers are quite comparable to those derived earlier in the quark model
and again show that the weak contribution to the electric dipole moments is
small and of order $10^{-32}e\cdot cm$. There are relatively few cancellations
involved in this calculation. We therefore do not expect very large higher
order corrections. The main uncertainties in the result come from the 
underlying assumptions: parity violating terms in the weak $|\Delta S|=1$ 
Lagrangian are negligible, parity violating terms in the strong light
quark mass sector
are of the same size as the parity conserving ones. 
The other source of uncertainty is the estimate of the parameter
$y_6(\mu )$ in the Wilson coefficient function of the effective four-quark 
operator $Q_6$, which we determined according to the renormalization 
scheme independent definition in \rcite{BURAS}.
This result should not be added to those obtained 
in the quark model. 
The value of $\langle\bar{q}q\rangle$ is related to the production of the 
constituent quark mass and the contribution as estimated here is thus
related to the one obtained in the quark model.

\section*{Acknowledgements}
We thank Eduardo de Rafael for suggesting this problem and collaborating
in the early stages of this work.
The work of E.P. is supported by EU Contract Nr. ERBCHBGCT 930442.

%\listoffigures
\end{document}